%% file: main.tex
\documentclass[runningheads]{llncs}
\usepackage{orcidlink}

\usepackage[dvipsnames]{xcolor} 

\hypersetup{
    colorlinks=true,       
    linkcolor=blue,        
    citecolor=green, 
    urlcolor=blue          
}

\usepackage[utf8]{inputenc}
\usepackage[T1]{fontenc}
\usepackage{graphicx}
\usepackage{amsmath,amssymb}
\usepackage{booktabs}
\usepackage{longtable}
\usepackage{xcolor}
\usepackage[ruled,vlined,linesnumbered]{algorithm2e}
\usepackage{tikz}
\usepackage{fontawesome5}
\usepackage{pgfplots}
\pgfplotsset{compat=1.17}
\usetikzlibrary{arrows.meta,positioning,calc}

\usepackage{array}
\makeatletter
\renewcommand\paragraph{\@startsection{paragraph}{4}{\z@}%
  {-6\p@ \@plus -1\p@ \@minus -1\p@}%
  {-0.5em \@plus -0.22em \@minus -0.1em}%
  {\normalfont\normalsize\itshape}}
\makeatother
\definecolor{tkblue}{HTML}{1F6FEB}
\definecolor{tknavy}{HTML}{0F2747}
\definecolor{tkgray}{HTML}{64748B}
\definecolor{tkamber}{HTML}{B45309}

\begin{document}
%
\title{EVISTA: Unified Multimodal Video Search with Evidence-Aware Temporal Path Optimization}
\titlerunning{EVISTA: Evidence-aware Video Indexing, Search, and Temporal Alignment}
%

\author{
Minh-Huy Tran$^{1}$\orcidlink{0009-0004-0535-5879} \and
Van-Thinh Vo$^{1}$\orcidlink{0009-0009-4815-1397} \and
Duy-Tan Nguyen$^{1}$\orcidlink{0009-0008-8732-7079} \and
Minh-Khoi Nguyen$^{1}$\orcidlink{0009-0000-0813-6576} \and
Anh-Quan Nguyen-Tran$^{1}$\orcidlink{0009-0000-3966-1795} \and
Loi Nguyen Khanh$^{\dag}$\orcidlink{0000-0003-4300-491X}
}


\authorrunning{M.H. Tran et al.}
%
\institute{
HCMUT EE Machine Learning \& IoT Lab, \\
Ho Chi Minh University of Technology, VNU-HCM, Vietnam, \\[2pt]
\texttt{\{huy.tranminh2005, 
thinh.vovan,  tan.nguyen2313054,  khoi.nguyenminh82005, quan.nguyenquan12,  nkloi\}@hcmut.edu.vn}}

\maketitle              
\begingroup
\renewcommand{\thefootnote}{\ensuremath{1}}
\footnotetext[1]{These authors contributed equally to this work as first authors.}
\renewcommand{\thefootnote}{\ensuremath{\dagger}}
\footnotetext[2]{Corresponding authors.}
\endgroup

\begin{abstract}
In the era of big data, precise event retrieval from large video collections has become increasingly important and challenging. In particular, locating multiple events from the same video in their temporal order is difficult when different events are described by different types of evidence, such as visual content, on-screen text, and spoken words. We develop EVISTA, a unified multimodal video retrieval system for the Ho Chi Minh City AI Challenge 2026, which integrates visual-semantic, OCR, ASR, object, and metadata search over a shared keyframe index, and propose TEMPO-P, an evidence-aware temporal path optimization method for ordered multi-event retrieval. Building EVISTA for this task reveals two limitations of independent event retrieval: a relevant video may be missed when one event is weakly retrieved, and independently retrieved event candidates may be temporally inconsistent or unable to represent events defined primarily by text or speech. TEMPO-P addresses these limitations by aggregating candidate videos across events and modalities, jointly scoring keyframes using visual, OCR, and ASR evidence, and selecting a compact temporally ordered path within each video through dynamic programming. On five organizer queries and 36 additional multi-event queries, TEMPO-P matches visual-only temporal alignment on visually described events (R@1 0.42 for both) and retrieves queries containing text-defined events, which visual-only alignment cannot represent (R@1 0.38). TEMPO-P is integrated into EVISTA as its ordered-event retrieval mode.

\end{abstract}
\input{secs/intro}
\input{secs/related-work}
\input{secs/method}
\input{secs/ui}
\input{secs/case-study}
\input{secs/conclusion}
\input{secs/ack}

%
%
%
\bibliographystyle{splncs04} 
\begingroup
\raggedright
\bibliography{refs}
\endgroup
%




\end{document}

%% file: secs/intro.tex
\section{Introduction}
Large video collections make locating a specific moment a major retrieval challenge~\cite{do2023news,lokovc2023interactive}. Users may recall visual content, on-screen text, speech, or objects, motivating multimodal evidence~\cite{heller2022multi}. Vision-language models enable free-text keyframe search~\cite{radford2021learning,lokovc2023interactive}, often complemented by Optical Character Recognition (OCR), Automatic Speech Recognition (ASR), and object detection~\cite{heller2022multi,amato2024visione}. However, offering separate retrieval modes or combining auxiliary cues with visual results does not by itself specify how textual evidence enters joint temporal alignment~\cite{lokovc2023interactive}. This distinction matters for ordered multi-event retrieval. In the \textit{Temporal Retrieval and Alignment of Key Events} (TRAKE) task of the Ho Chi Minh City AI Challenge, a query describes successive moments, such as cooking steps, and requires a video and one representative keyframe per event in the correct order~\cite{do2025toward}. Independent retrieval may select frames from different videos or in the wrong order, motivating temporal query processing~\cite{heller2022multi}. Visual-only alignment also cannot directly represent events defined exclusively by an on-screen title or spoken phrase.
\\
\indent We propose \textbf{TEMPO-P} to jointly integrate multimodal evidence and temporal alignment. It retrieves event sequences whose keyframes belong to one video and occur in strictly increasing order. OCR/ASR evidence expands the candidate-video set and enters event scoring, including for text-only events. A layered monotonic dynamic program jointly aligns events over candidate videos' indexed keyframes and produces video-ranking scores, without decoding raw video at query time. 
\\
\indent Then, we integrate TEMPO-P into \textbf{EVISTA} (Evidence-aware Video Indexing, Search, and Temporal Alignment), an interactive multimodal video retrieval system supporting both single-event and ordered multi-event queries over shared indices. Single-event queries combine Vietnamese or English free-text descriptions, encoded by a multilingual multimodal model, with OCR, ASR, object, and metadata conditions. For ordered multi-event queries, TEMPO-P aligns events described by visual content, OCR/ASR cues, or both. In summary, this paper makes the following contributions:
\begin{itemize}
  \item \textbf{An unified multimodal retrieval system.} EVISTA combines multilingual semantic, OCR, ASR, object, and metadata search within a single query.
  \item \textbf{TEMPO-P for ordered multi-event retrieval.} A layered monotonic dynamic program enforces video consistency and temporal order while incorporating OCR/ASR evidence, including for events described only by text or speech.
\end{itemize}

%% file: secs/related-work.tex
\section{Related Work}
\subsection{Video Retrieval}
Modern video retrieval systems use vision-language models to support natural-language
search over large collections. CLIP~\cite{radford2021learning} and subsequent joint
embedding models map visual content and text into a shared space, enabling semantic
retrieval from free-text queries, and interactive systems such as VISIONE
5.0~\cite{amato2024visione} combines semantic retrieval with object, color, and metadata
search. These systems show the effectiveness of multimodal indexing for interactive
retrieval, but primarily address the retrieval of individual moments rather than the
joint alignment of several temporally ordered events.
\subsection{Multimodal Search}

Multimodal video retrieval extends semantic search with complementary signals such as
OCR, ASR, object annotations, and metadata, which help when the target moment is
characterized by information that is hard to capture by visual similarity alone. Vo et
al.~\cite{vo2025enhanced} combine CLIP and BEiT-3 embeddings with object, color, and OCR
information for cross-modal event retrieval, and other systems explore LLM-based query
expansion~\cite{ma2024leveraging}. EVISTA follows this paradigm and additionally uses
ASR as a retrieval signal. More importantly, TEMPO-P uses visual, OCR, and ASR evidence
inside temporal alignment rather than only as independent retrieval channels or
post-processing filters.
\subsection{Temporal Retrieval}
Temporal retrieval requires several events to satisfy temporal relations. Earlier
interactive systems combine independently retrieved results under temporal
constraints~\cite{heller2022multi}, and Vo et al.~\cite{vo2025enhanced} extend this to
cross-modal events followed by temporal re-ranking. More recent TRAKE-oriented methods
formulate the problem as sequence alignment: VidAlign~\cite{nguyen2025vidalign} fuses
event scores by temporal dynamic programming (TDP-Fuse), DANTE~\cite{luu2025integrated}
applies dynamic programming over keyframe similarities with a temporal penalty, and
MADTempo~\cite{vu2027madtempo} aggregates scores across sequential video segments with
query augmentation. Related moment-retrieval methods use temporal
re-ranking~\cite{tran2025towards} and adaptive bidirectional temporal
search~\cite{nguyen2025lightweight}. These alignment methods score events by visual--semantic similarity. An event whose
defining evidence is an on-screen text or a spoken phrase, therefore has no direct
representation in the alignment objective, and when candidates are retrieved per event
before alignment, one weak event can discard a relevant video. TEMPO-P addresses both by
combining video-level candidate aggregation with an evidence-aware event score that
enters a monotonic temporal path optimization. 

%% file: secs/method.tex
\section{Methodology}
\label{sec:method}
\subsection{EVISTA: System Overview and Indexed Representations}
\label{sec:repr}
\begin{figure}[h]
\centering
\includegraphics[width=\textwidth]{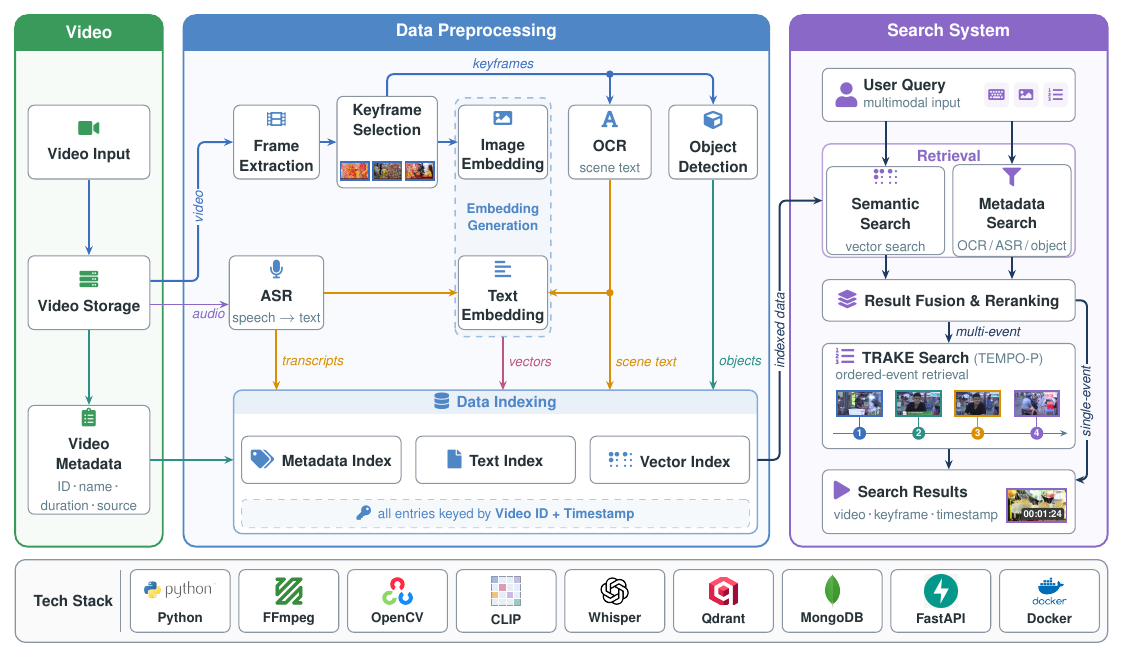}
\caption{Overview of EVISTA. Keyframes are extracted from the input videos and
represented by visual and caption embeddings, OCR and ASR text, and object annotations,
which are stored in a vector database and a metadata database, and shared by single-event
search and TEMPO-P.}
\label{fig:system-overview}
\end{figure}

\paragraph{\textbf{System Overview.}} EVISTA indexes keyframes together with visual, caption, OCR, ASR, object and video
metadata (Fig.~\ref{fig:system-overview}). Single-event queries are answered by the
retrieval channels of Sec.~\ref{sec:single}; ordered multi-event (TRAKE) queries invoke
TEMPO-P (Sec.~\ref{sec:tempop}), which reuses the same indices. Results of both modes are
returned as ranked keyframes or event tuples through the interface of Sec.~\ref{sec:ui}.

\paragraph{\textbf{Data Processing and Indexing.}} We use \emph{keyframe} for an indexed frame, the retrieval unit, regardless of how it was
sampled. For videos with scene transitions, keyframes are selected by the adaptive
segmentation of our previous system~\cite{vo2025enhanced}; the stationary-camera traffic
videos of AIC's dataset 2026, which contain no transitions, are sampled uniformly at one-second
intervals. Each keyframe is identified by its video and its frame index in the original video, and carries the evidence listed in Table~\ref{tab:repr}. Vectors are stored in
Qdrant and metadata in MongoDB Atlas. TEMPO-P uses the visual, OCR, and ASR evidence;
object and metadata information serve single-event filtering.

\begin{table}[h]
\centering\scriptsize
\setlength{\tabcolsep}{4pt}
\caption{Evidence attached to each keyframe.}
\label{tab:repr}
\begin{tabular}{@{}lll@{}}
\toprule
Evidence & Representation & Used for\\
\midrule
Visual   & BEiT-3~\cite{wang2023image}, PE-CLIP~\cite{bolya2025PerceptionEncoder}, jina-embeddings-v5-omni~\cite{honicke2026jina} & semantic search; TEMPO-P\\
Caption  & Qwen3-VL-4B-Instruct~\cite{yang2025qwen3} captions, text-encoded & semantic search\\
OCR      & EasyOCR~\cite{easyocr2024} detection, VietOCR~\cite{vietocr2020} recognition & OCR search; TEMPO-P\\
ASR      & PhoWhisper-small~\cite{PhoWhisper} segments with frame ranges & ASR search; TEMPO-P\\
Object   & YOLO11n~\cite{yolo11_ultralytics} detections (traffic videos) & object filtering\\
Metadata & video id, frame index, timestamp, fps & indexing; submission\\
\bottomrule
\end{tabular}
\end{table}

\subsection{Single-Event Search}
\label{sec:single}

\begin{figure}[h]
\centering
\includegraphics[width=\linewidth]{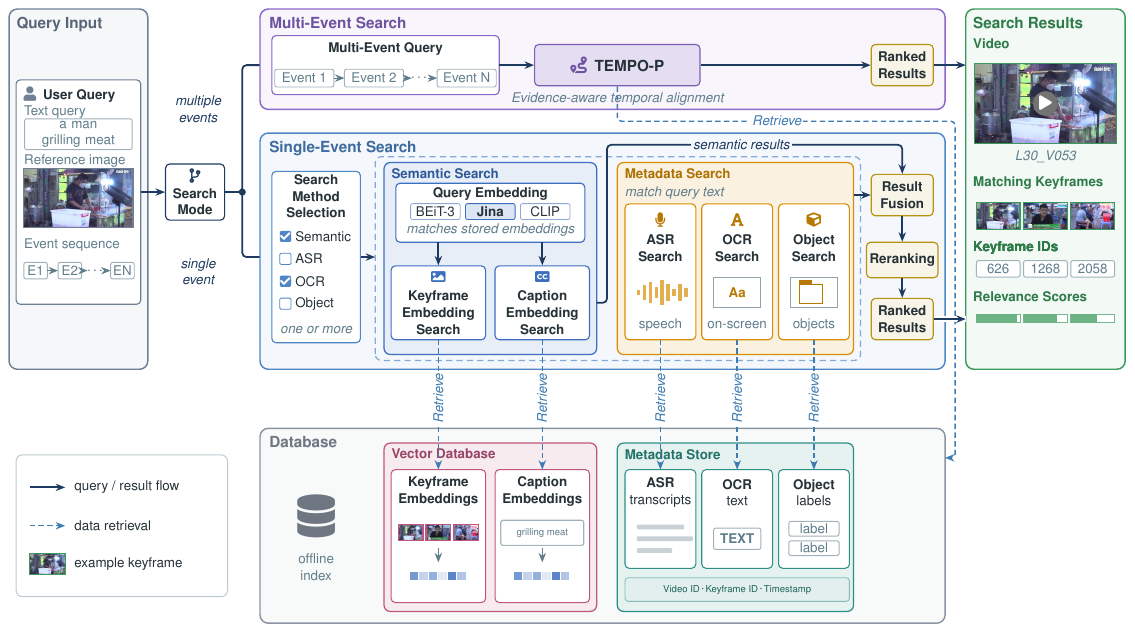}
\caption{Retrieval modes of EVISTA. Single-event search combines semantic retrieval over
keyframe and caption embeddings with OCR-, ASR- and object-based search; ordered
multi-event search invokes TEMPO-P. Both modes share the indexed representations.}
\label{fig:se}
\end{figure}

\paragraph{\textbf{Notation and matching functions.}} Table~\ref{tab:notation} defines
the notation and the matching functions shared by single-event search and TEMPO-P. OCR
and ASR cues are matched by fuzzy full-text search, which tolerates recognition errors
such as missing diacritics; the evidence weights $w$ are fixed design choices encoding
verbatim $>$ fuzzy $>$ none and were not tuned.

{\scriptsize
\setlength{\tabcolsep}{4pt}
\renewcommand{\arraystretch}{1.15}
\setlength{\LTpre}{6pt}\setlength{\LTpost}{6pt}
\begin{longtable}{@{}p{0.27\linewidth}p{0.69\linewidth}@{}}
\caption{Notation and matching functions shared by single-event search and TEMPO-P.}
\label{tab:notation}\\
\toprule
Symbol & Definition\\
\midrule
\endfirsthead
\multicolumn{2}{@{}l}{\emph{Table~\ref{tab:notation} (continued)}}\\
\toprule
Symbol & Definition\\
\midrule
\endhead
\bottomrule
\endlastfoot
$V$ & a video\\
$f^{(1)}<\dots<f^{(T_V)}$ & the $T_V$ keyframes of $V$ in temporal order, each identified by its frame index in $V$\\
$\iota(f^{(t)})$ & ordinal index of a keyframe, $\iota(f^{(t)})=t$\\
$q=(q^{\mathrm{v}},q^{\mathrm{o}},q^{\mathrm{a}})$ & a query, or one event of a TRAKE query: visual description, on-screen text cue, speech cue (any may be empty)\\
$\phi$ & embedding model mapping text and keyframes into a shared space (a visual model of Table~\ref{tab:repr})\\
$s(q,f)$ & semantic score, $s(q,f)=\cos\big(\phi(q^{\mathrm{v}}),\phi(f)\big)$\\
$\mathrm{ANN}(q,K)$ & the $K$ keyframes with the highest $s(q,f)$\\
$\mathcal{M}^{\mathrm{o}}(q)$ & keyframes whose OCR text matches $q^{\mathrm{o}}$\\
$\mathcal{M}^{\mathrm{a}}(q)$ & keyframes inside a speech segment $[\tau^{s},\tau^{e}]$ (frame indices) whose transcript matches $q^{\mathrm{a}}$\\
$w$ & match weight: $1$ for a verbatim match, $\max(1-0.02r,\,0.5)$ for the $r$-th fuzzy OCR match, $0.9$ for a fuzzy ASR match\\
$e^{\mathrm{o}}(q,f)$ & graded OCR evidence, $\max w(g)$ over $g\in\mathcal{M}^{\mathrm{o}}(q)$ with $|f-g|\le5$ video frames ($0$ if none)\\
$e^{\mathrm{a}}(q,f)$ & graded ASR evidence, $\max w$ over matching segments $[\tau^{s},\tau^{e}]\ni f$ ($0$ if none)\\
$e(q,f)$ & textual evidence, $e=\max(e^{\mathrm{o}},e^{\mathrm{a}})\in[0,1]$\\
\end{longtable}}

\paragraph{\textbf{Standalone search.}} As in typical interactive systems~\cite{vo2025enhanced,amato2024visione,do2025toward}, each channel in Fig.~\ref{fig:se} can retrieve results independently: semantic search returns $\mathrm{ANN}(q,K)$, OCR and ASR search return $\mathcal{M}^{\mathrm{o}}(q)$ and $\mathcal{M}^{\mathrm{a}}(q)$ over all videos ranked by text relevance, and object search filters keyframes by detection metadata.

\paragraph{\textbf{Hybrid search.}} As a supporting interface feature, OCR and ASR cues can filter semantic results when a user recalls a word shown on screen or spoken. Semantic retrieval produces $\mathcal{L}=\mathrm{ANN}(q,K)$, sorted by decreasing $s(q,f)$, and the filtered result is $\mathcal{R}=[f\in\mathcal{L}:f\in\mathcal{M}^{\mathrm{o}}(q)\land f\in\mathcal{M}^{\mathrm{a}}(q)]$, where an empty cue imposes no condition. The order of $\mathcal{L}$ is preserved: the semantic score remains the only ranking score, and the cues only determine which keyframes remain. The matching is restricted to the videos in $\mathcal{L}$, with one text-index lookup per distinct video for each active cue.

\subsection{TEMPO-P for Ordered Multi-Event Retrieval}
\label{sec:tempop}

\begin{figure}[h]
\centering
\includegraphics[width=\textwidth]{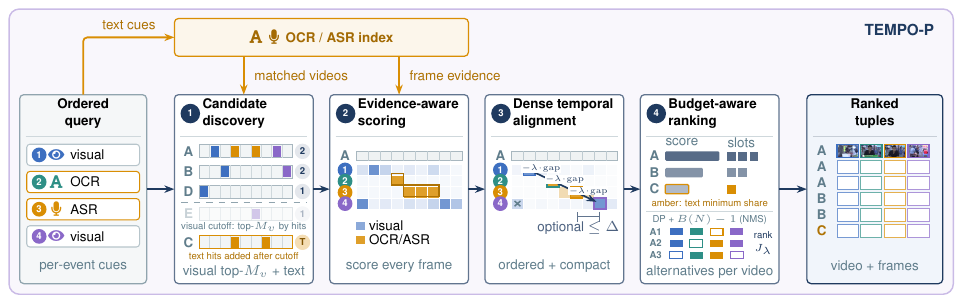}
\caption{TEMPO-P pipeline. Candidate videos come from per-event visual hits and OCR/ASR
matches (1). Every keyframe of a candidate video is scored for every event, with OCR/ASR
evidence (amber) added to the visual score and alone defining events described only by
text (2). A DP selects one keyframe per event in temporal order, penalizing temporal
spread and optionally bounded by a maximum gap (3). The submission budget is shared among
the best videos, each contributing its optimal tuple and ranked alternatives (4).}
\label{fig:search_tempo_p}
\end{figure}

A TRAKE query describes several events in temporal order and asks for a video and one
keyframe per event~\cite{do2025toward}. TEMPO-P (Temporal Evidence-aware Multimodal Path
Optimization with Probabilistic component selection) formulates this as a path
optimization problem over the ordered query $Q=\langle q_1,\dots,q_N\rangle$, each event
$q_i=(q_i^{\mathrm{v}},q_i^{\mathrm{o}},q_i^{\mathrm{a}})$ using the matching functions of
Table~\ref{tab:notation}: every event is scored from visual and OCR/ASR
evidence, so that textual evidence can \emph{define} an event, and the selected keyframes
are constrained to be temporally ordered and compact. It proceeds in four stages
(Fig.~\ref{fig:search_tempo_p}); the ``probabilistic component selection'' refers only to
the allocation of the submission budget in the last stage.

\paragraph{\textbf{Problem Fomulation.}} For a video $V$, TEMPO-P seeks
\begin{equation}
\label{eq:problem}
F^\star_V=\operatorname*{arg\,max}_{F}\ J_\lambda(F)\quad\text{s.t.}\quad
f_1<\dots<f_N,\quad f_{i+1}-f_i\le\Delta,
\end{equation}
and returns videos ranked by $J^\star_\lambda(V)=J_\lambda(F^\star_V)$. Ordering and
$\Delta$ is expressed in video frames, from which the submitted answers are derived.

\paragraph{\textbf{Stage 1: candidate video construction.}} Let $\mathrm{vid}(\cdot)$ map a set
of keyframes to the videos containing them, and let $\mathcal{T}_i$ be the first
$K_{\mathrm{txt}}=110$ keyframes of $\mathcal{M}^{\mathrm{o}}(q_i)\cup\mathcal{M}^{\mathrm{a}}(q_i)$
(80 verbatim and 30 fuzzy matches). Then
\begin{equation}
\label{eq:cand}
\mathcal{V}=\mathrm{Top}_{M_v}\Big(\bigcup_{i=1}^{N}\mathrm{vid}\big(\mathrm{ANN}(q_i,M)\big)\Big)\ \cup\ \bigcup_{i=1}^{N}\mathrm{vid}(\mathcal{T}_i).
\end{equation} Candidate videos are formed by union rather than intersection,
so that a weakly retrieved event cannot eliminate an otherwise relevant video; $M$ is
the per-event retrieval depth, and the visual union is ranked by the number of events
hitting each video before being truncated to $M_v$ videos.

\paragraph{\textbf{Stage 2: evidence-aware event scoring.}} For every keyframe $f$ of every
$V\in\mathcal{V}$,
\begin{equation}
\label{eq:score}
\sigma_i(f)=s(q_i,f)+\beta\,e(q_i,f),
\end{equation}
with $\phi$ instantiated by jina-embeddings-v5-omni~\cite{honicke2026jina}. For an
\emph{evidence-only} event ($q_i^{\mathrm{v}}$ empty) the semantic term is omitted,
$\sigma_i(f)=\beta\,e(q_i,f)$, which visual-only alignment cannot represent. OCR/ASR
evidence thus enters the objective itself rather than a late fusion of ranked
lists~\cite{cormack2009reciprocal}.

\paragraph{\textbf{Stage 3: temporal path optimization.}} A tuple is scored by
\begin{equation}
\label{eq:obj}
J_\lambda(F)=\sum_{i=1}^{N}\sigma_i(f_i)-\lambda\sum_{i=1}^{N-1}\big(\iota(f_{i+1})-\iota(f_i)\big),
\end{equation}
whose second term prefers compact alignments over paths assembled from visually similar
but distant shots. Spread is measured in keyframe steps because keyframes are placed at
shot changes, so the ordinal distance counts intervening shots; all TRAKE videos share
this sampling, so $\lambda$ has the same meaning across videos. Problem~\eqref{eq:problem}
is solved exactly over all $T_V$ keyframes of $V$ (at most 4000) by a layered DP, in the
spirit of dynamic time warping~\cite{sakoe1978dynamic} and DP-based sequence alignment in
video~\cite{zhukov2019cross,dvornik2021drop,han2022temporal}:
\begin{align}
D_1(t)&=\sigma_1\big(f^{(t)}\big),\label{eq:dp0}\\
D_i(t)&=\sigma_i\big(f^{(t)}\big)+\max_{t'<t,\ f^{(t)}-f^{(t')}\le\Delta}\big(D_{i-1}(t')-\lambda(t-t')\big),\label{eq:dp}\\
J^\star_\lambda(V)&=\max_t D_N(t),\qquad F^\star_V=\mathrm{backtrack}(D_N).\label{eq:dpend}
\end{align}

\paragraph{\textbf{Stage 4: budget-aware ranking.}} Stages 1--3 are deterministic; the
probabilistic component of TEMPO-P applies only here, to budget allocation under the
submission constraint. A submission holds at most $C_{\max}$ tuples, and spreading them
over several plausible videos rather than one keeps the correct video in the submission
when its $J^\star_\lambda$ is not the highest. The top-$K_v$ videos by $J^\star_\lambda$
share the slots in proportion to
$\pi(V)\propto\exp J^\star_\lambda(V)$, with a minimum share for videos endorsed by
OCR/ASR evidence. Within a video, $\mathcal{C}_i(V)$ holds the DP keyframe of event $q_i$
and the next $B(N)-1$ keyframes by $\sigma_i$ after temporal non-maximum suppression, and
the valid tuples of $\prod_i\mathcal{C}_i(V)$ are emitted in decreasing $J_\lambda$. This
normalization only allocates the budget and adds no ranking signal beyond $J_\lambda$.

\paragraph{\textbf{Complexity.}} Computing $\sigma$ costs $O(N\,T_{\mathrm{cand}}\,d)$ for the
$T_{\mathrm{cand}}=\sum_{V\in\mathcal{V}}T_V$ candidate keyframes ($d$ the embedding
dimension). A naive evaluation of Eq.~\eqref{eq:dp} costs $O(N\,T_V^2)$ per video; since
$-\lambda(t-t')=\lambda t'-\lambda t$, the inner maximum is a sliding-window maximum of
$D_{i-1}(t')+\lambda t'$, maintained with a monotone deque, which gives
$O(N\,T_{\mathrm{cand}})$ in total. Stage~4 enumerates at most $B(N)^N$ tuples per kept
video ($B(N)=10,4,3,3,2$ for $N=2,\dots,6$, i.e.\ 64--243 tuples), which is small
compared with scoring.


%% file: secs/ui.tex
\section{User Interface}
\label{sec:ui}
\begin{figure}[h!]
\centering
\fbox{\includegraphics[width=0.75\linewidth]{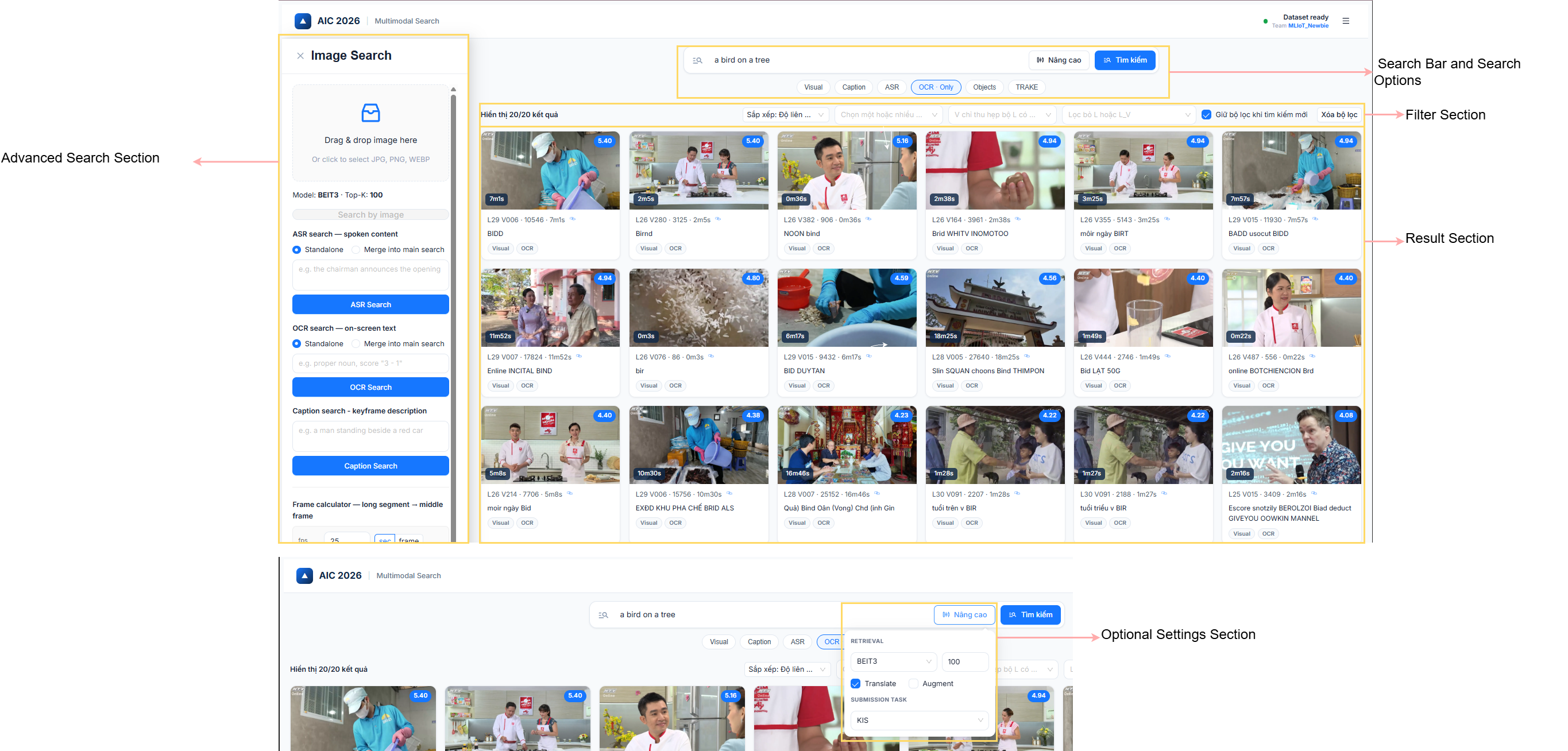}}\\[1pt]
{\scriptsize (a) Unified query interface and result grid}\\[4pt]
\fbox{\includegraphics[width=0.75\linewidth]{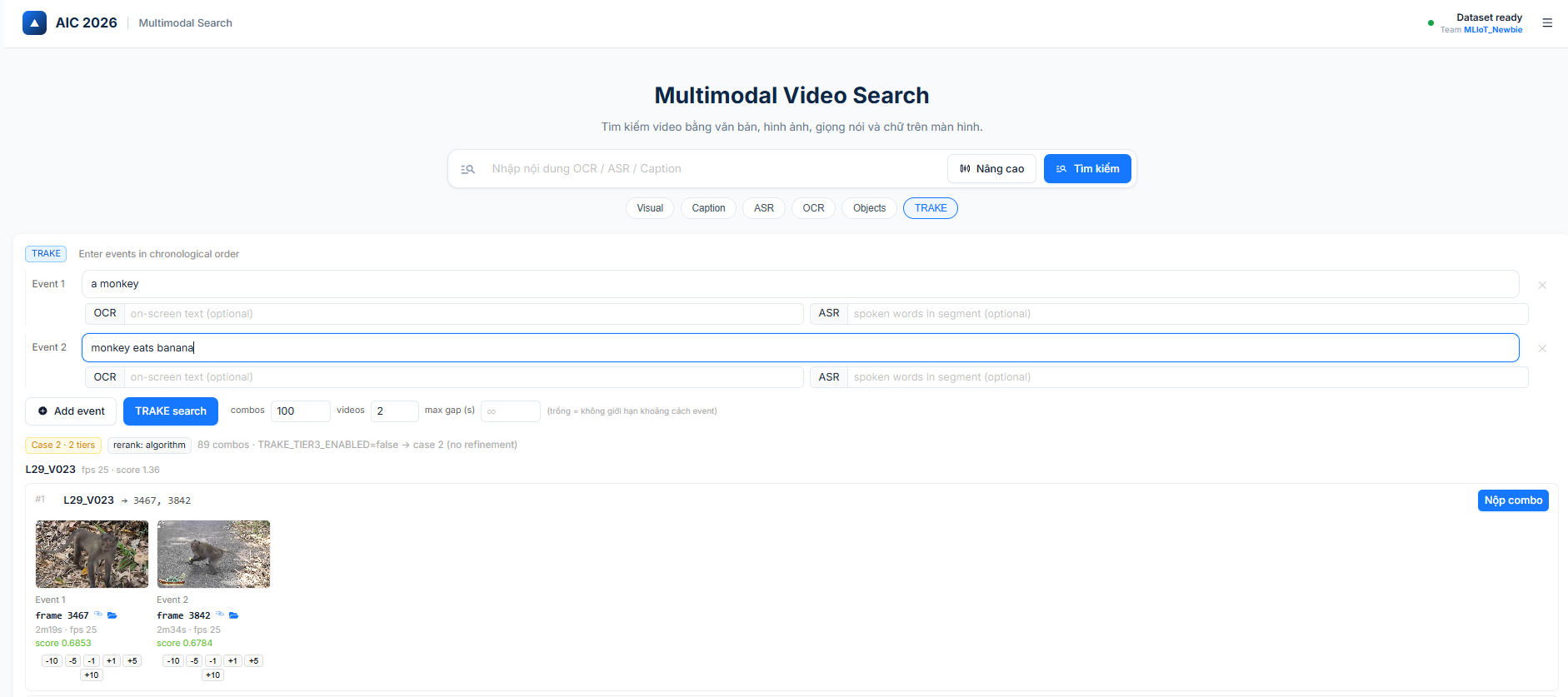}}\\[1pt]
{\scriptsize (b) TRAKE query construction and tuple review}
\caption{User interface of EVISTA. (a) A central query box with modality chips and a
ranked keyframe grid with visible hybrid and object constraints. (b) An ordered event
sequence with visual, OCR or ASR cues and the returned keyframe tuples for review and
submission.}
\label{fig:ui}\label{fig:ui-overview}\label{fig:ui-trake}
\end{figure}

The interface supports interactive retrieval through three complementary functions: (i) \textit{Unified query construction.} A central query box provides modality chips for visual, caption, ASR, OCR, object, and TRAKE search, allowing users to switch retrieval pipelines without changing pages (Fig.~\ref{fig:ui}(a)). 
Advanced options configure the embedding model, result depth, query translation/augmentation, and submission task. OCR and ASR support standalone and hybrid retrieval, with active constraints displayed alongside the results. (ii) \textit{Result browsing and inspection.} Compact result cards show keyframe thumbnails, video/frame identifiers, timestamps, and retrieval modalities; visual-search scores are the cosine similarities returned by Qdrant. Dataset/video filtering, relevance/time sorting, and pagination support browsing. Selecting a card opens the Video Result Inspector, which joins visual, caption, OCR, ASR, and object information using $(\mathit{video\_id},\mathit{frame\_id})$. The inspector displays frame-level metadata and detected objects, highlights matching OCR/ASR spans as qualitative evidence rather than per-modality relevance scores, and provides access to nearby keyframes, video playback, and submission. Frequently used controls remain near the results, while detailed evidence is available on demand. Finally, (iii) \textit{Temporal query construction and tuple review.} The TRAKE interface presents a natural-language description as an editable, ordered event sequence (Fig.~\ref{fig:ui}(b)). Users specify each event through a visual description, OCR cue, or ASR cue, making temporal order explicit and accommodating events identifiable through on-screen text or speech. Returned candidates appear as temporally ordered keyframe tuples for review and submission.

%% file: secs/case-study.tex
\section{Experiments and Case Studies}
After the Pre-Final Round in the AI Challenge Ho Chi Minh City 2026, our system achieved 78/86 points on average, efficiently retrieving both per-scene and temporal events queries. Specifically, our search strategies are in the following sections. 
\subsection{Single-Event Retrieval}

\begin{figure}[t]
    \centering
    \fbox{\includegraphics[width=0.9\linewidth]{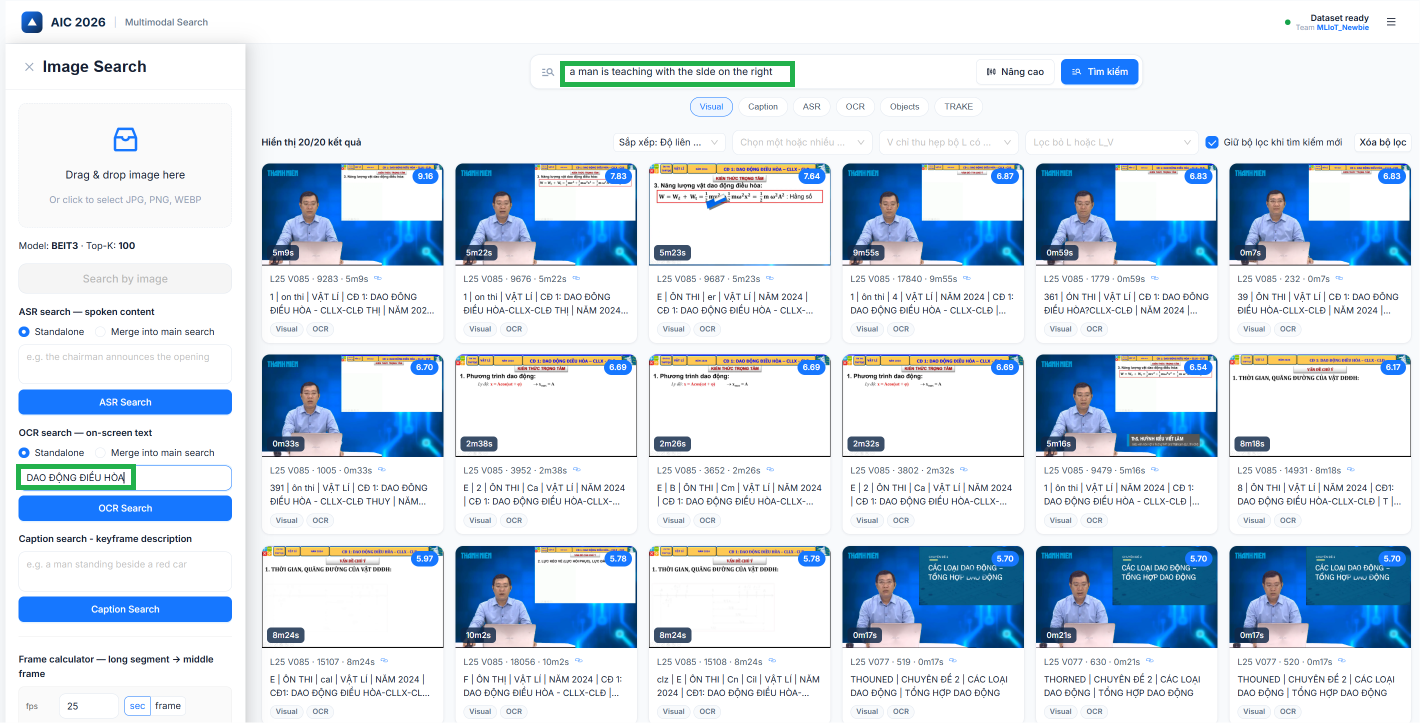}}\\[1pt]
    {\scriptsize (a) OCR cue applied as a hybrid filter on semantic results}\\[4pt]
    \fbox{\includegraphics[width=0.9\linewidth]{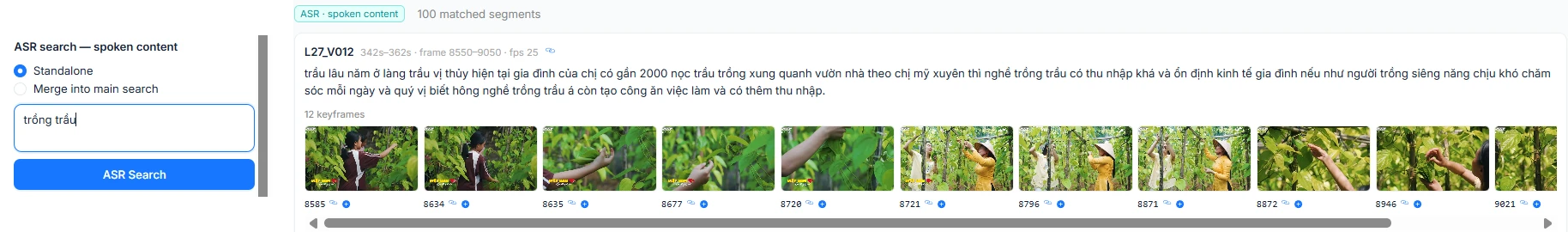}}\\[1pt]
    {\scriptsize (b) ASR-only search: the answer segment is the first match}
    \caption{Single-event queries resolved by textual evidence. (a) The on-screen phrase
    ``\textit{GIAO DONG DIEU HOA}'' is used as an OCR filter instead of semantic text.
    (b) The answer to a QA query is found only in speech: ASR searches for
    ``\textit{trong trau}'' returns the segment of video L27\_V012 containing the answer
    keyframe 8721.}
    \label{fig:hybrid}\label{fig:asr}
\end{figure}

The \textbf{semantic search} and \textbf{metadata-based search} are similar to our previous system \cite{vo2025enhanced}, with an appropriate query for each modality to return the top candidate results.
\paragraph{\textbf{Text and speech evidence.}} A single modality is often insufficient for a
query (Fig.~\ref{fig:hybrid}). \emph{OCR (hybrid).} In Fig.~\ref{fig:hybrid}a the query asks
for a teacher lecturing on ``\textit{GIAO DONG DIEU HOA}'' (harmonic oscillation). Adding
this phrase to the semantic query does not help, since it is written on the slide rather
than depicted; instead, it is used as an OCR cue that filters the semantic results.
\emph{ASR only.} The QA query in Fig.~\ref{fig:hybrid}b describes a report on a
traditional craft village in the Mekong Delta whose product is a characteristic food of
southern Vietnam, filmed in the betel village of "Vi Thuy", and asks for the city the
village belongs to today. Semantic search with the full description does not return the
answer moment among its top-1000 keyframes (neither with jina-embeddings-v5-omni ~\cite{honicke2026jina} nor with
BEiT-3~\cite{wang2023image}), and OCR search for the village name finds the right video, but not the moment.
The answer is only spoken: ASR search for ``\textit{trong trau}'' (growing betel)
returns as its first segment the passage (342--362\,s, frames 8550--9050) in which the
speaker introduces the betel growers of Vi Thuy, from which the answer, CANTHO, follows.

\subsection{Ordered-Event Retrieval with TEMPO-P}
\label{sec:cs-trake}

\noindent\textbf{Setup.} Following the TRAKE task definition~\cite{do2025toward}, we
analyse five TRAKE queries for which reference answers
(target video and per-event frame ids or ranges) were made available by the
organizers: four described visually (Q1--Q4, 3--4 events each) and one whose second
event is identified by an on-screen caption (Q5). These are not the official
final-round evaluation. With five queries, this part is a case study; we report what each mechanism enables rather than claim statistical differences.
A tuple counts as correct if it is in the target video and every event lies within
$\pm30$ frames ($\approx$1.2\,s) of its annotation. Keyframes are indexed with the
jina-embeddings-v5-omni model in Qdrant (about
$9.2\times10^5$ keyframes); $M=300$, $M_v=40$, $K_v=5$,
$C_{\max}=100$, $\lambda=0.005$, $\beta=0.3$, no maximum gap $\Delta$.

\begin{figure}[t]
\centering\scriptsize
\setlength{\tabcolsep}{3pt}
\begin{tabular}{@{}p{0.11\linewidth}p{0.85\linewidth}@{}}
\toprule
Query & Events (cue type)\\
\midrule
Q5 (Case 1) & $E_1$ a man in black grilling meat with two pairs of tongs (visual);
$E_2$ caption ``\textit{com tam Thanh Nien}'' (\textbf{OCR only}, no visual description);
$E_3$ two customers on a motorbike waiting to buy rice (visual)\\
Q2 (Case 2) & $E_1$ a red star with a yellow border inside a silver ring; $E_2$ the female
host holding a string hung with small objects; $E_3$ a middle-aged man in a black shirt
with an orange pattern holding an object; $E_4$ a toy panda with a glowing belly hanging
from a string (all visual)\\
Q1--Q4 (Case 3) & 3--4 visual events each, e.g.\ Q1: a chef brushes oil onto the pan;
places eel on the grill pan; presses it with a wooden spatula; the first piece is put on
a plate\\
\bottomrule
\end{tabular}\\[4pt]
\setlength{\tabcolsep}{1pt}
\begin{tabular}{cccccc}
\includegraphics[width=.163\linewidth]{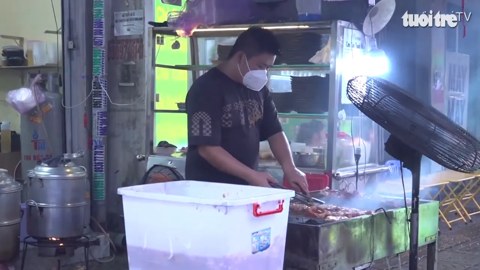} &
\includegraphics[width=.163\linewidth]{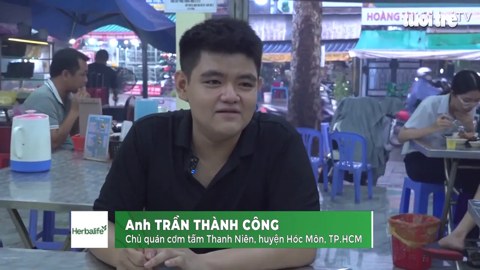} &
\includegraphics[width=.163\linewidth]{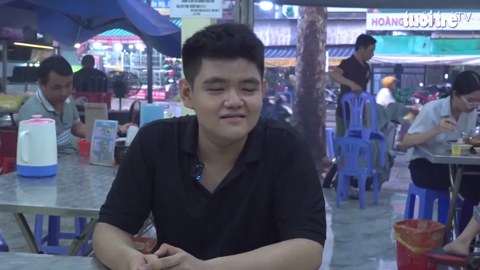} &
\includegraphics[width=.163\linewidth]{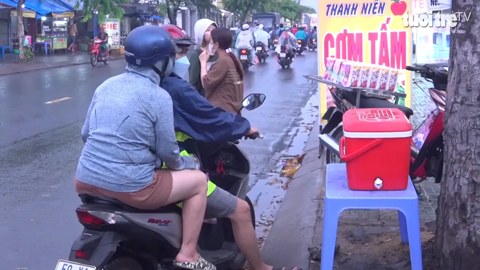} &
\includegraphics[width=.163\linewidth]{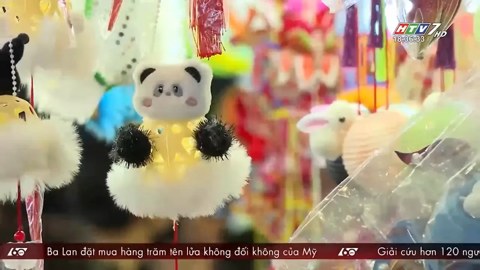} &
\includegraphics[width=.163\linewidth]{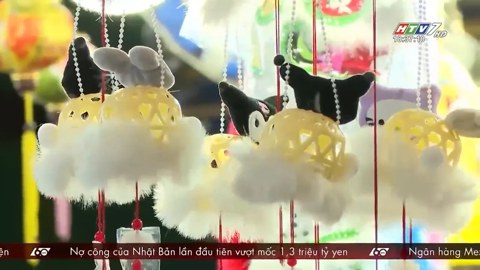}\\
(a) Q5 $E_1$, 626 & (b) Q5 $E_2$, 1268 & (c) visual $E_2$, 1457 & (d) Q5 $E_3$, 2058 &
(e) Q2 $E_4$, 9758 & (f) $\lambda{=}0$, 10673
\end{tabular}
\caption{Case-study queries (top) and keyframes with frame indices (bottom). Q5: The second event is defined by the on-screen caption in (b); a visual description of it
selects (c), later in the same interview, and without the caption. Q2: the annotated last
event (e) and the look-alike shot (f) selected without the proximity penalty.}
\label{fig:cases}\label{fig:q5}\label{fig:q2}
\end{figure}

\paragraph{\textbf{Case 1: textual evidence can define an event.}} In Q5 (Fig.~\ref{fig:cases}),
$E_2$ is the interview segment introducing the stall owner, identifiable by the caption
naming the stall (b). Written as a visual description, $E_2$ is aligned to a later frame
of the same interview (c), 189 frames after the annotated moment, and without the
caption. As an \emph{evidence-only} event, $\sigma_2=\beta\,e(q_2,f)$ is non-zero only
around the caption, the DP is forced through it, and TEMPO-P returns the correct tuple at
rank~3; requiring every event in the top-$M$ of one video finds it only at rank~22, and a visual-only DP, including DANTE, cannot represent the query at all.

\paragraph{\textbf{Case 2: temporal coherence filters look-alike shots.}} In Q2, without the
proximity penalty ($\lambda=0$) the best-scoring tuple takes $E_4$ from a shot of
hanging plush lanterns about 900 frames later (f) instead of the annotated panda (e), and
the correct tuple drops from rank~1 to rank~4; with the penalty, the compact, correct
alignment is ranked first.

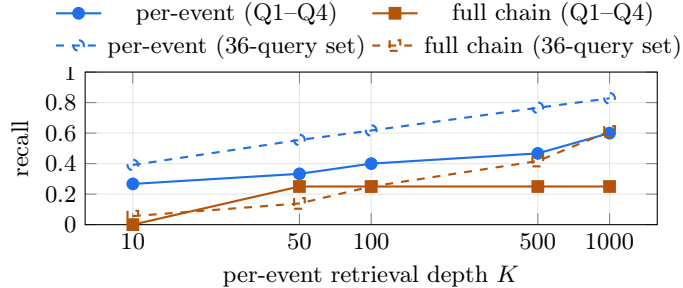
\begin{figure}[h]
\centering
\begin{tikzpicture}
\begin{axis}[width=0.75\linewidth,height=3.6cm,xmode=log,log basis x=10,
  xtick={10,50,100,500,1000},xticklabels={10,50,100,500,1000},
  xlabel={per-event retrieval depth $K$},ylabel={recall},ymin=0,ymax=1,
  legend columns=2,legend style={font=\footnotesize,at={(0.5,1.03)},anchor=south,draw=none},
  grid=both,grid style={gray!20},
  tick label style={font=\footnotesize},label style={font=\footnotesize}]
\addplot[tkblue,thick,mark=*] coordinates
  {(10,0.267)(50,0.333)(100,0.4)(500,0.467)(1000,0.6)};
\addplot[tkamber,thick,mark=square*] coordinates
  {(10,0.0)(50,0.25)(100,0.25)(500,0.25)(1000,0.25)};
\addplot[tkblue,thick,dashed,mark=o] coordinates
  {(10,0.391)(50,0.555)(100,0.617)(500,0.766)(1000,0.828)};
\addplot[tkamber,thick,dashed,mark=square] coordinates
  {(10,0.056)(50,0.139)(100,0.25)(500,0.417)(1000,0.611)};
\legend{per-event (Q1--Q4), full chain (Q1--Q4),
        per-event (36-query set), full chain (36-query set)}
\end{axis}
\end{tikzpicture}
\caption{Deeper per-event retrieval raises per-event recall much faster than the recall
of complete event chains, on Q1--Q4 (solid) and on the 36-query set of
Table~\ref{tab:set36} (dashed), motivating video-level candidate aggregation before
temporal grounding.}
\label{fig:recall}
\end{figure}

\paragraph{\textbf{Case 3: candidate discovery must precede temporal reasoning.}} For
Q1--Q4, joint-embedding search is effective for single shots~\cite{lokovc2023interactive},
yet increasing the per-event retrieval depth raises per-event recall but not the
recall of complete event chains (Fig.~\ref{fig:recall}): an abstract event such as
``brushing oil onto the pan'' never enters the top-1000, and a single missing event
removes the video. Taking the union of candidate videos and grounding every event
densely within them removes these candidate misses; the remaining failures (Q1, Q4)
are events with low visual similarity (an action on a pan; jersey numbers defined
arithmetically), i.e., \ semantic rather than temporal errors. When the target video
is returned, localization is accurate (mean error 2.5 frames, 62.5\% of events exact).

\paragraph{\textbf{Extended evaluation set.}} 
To test these observations beyond five queries, we
authored 36 TRAKE queries on 36 further videos (news, cooking, travel, lecture, and
cycling broadcasts; traffic-camera videos excluded, as in the challenge). Each query has
four events inside one contiguous segment (median 37\,s, at most 120\,s), written in the
organisers' style after inspecting every keyframe of the segment, with the frame range of
each event's shot as reference answer. 13 queries contain an event identified only by
on-screen text, 11 add an OCR cue to visual descriptions, and 12 are purely visual.
Because the author saw the keyframes and every OCR cue was checked to be recognized
inside its reference range, this set is an upper-bound feasibility study: it measures
what multimodal alignment can achieve when the textual evidence is correctly recognized,
not its robustness to OCR errors.
\begin{table}[t]
\centering\scriptsize
\setlength{\tabcolsep}{3pt}
\caption{Diagnostics. Organiser queries: R@1 on Q1--Q4 and rank of the correct tuple on
Q5. Self-authored set: R@1 by query type (Vis.: visual only; +OCR: OCR cue added to
visual descriptions; Text: contains an event identified only by on-screen text) and R@10
over all 36 queries. Defaults: $\beta=0.3$, $\lambda=0.005$. n.e.: not expressible;
--: not run.}
\label{tab:diag}\label{tab:set36}\label{tab:set36x}
\begin{tabular}{@{}lccccccc@{}}
\toprule
 & \multicolumn{2}{c}{Organiser} & \multicolumn{5}{c}{Self-authored set}\\
\cmidrule(lr){2-3}\cmidrule(l){4-8}
Variant & Q1--Q4 & Q5 & Vis. & +OCR & Text & All & R@10\\
\midrule
\textbf{TEMPO-P}                          & \textbf{0.50} & \textbf{3} & \textbf{0.42} & 0.27 & \textbf{0.38} & \textbf{0.36} & \textbf{0.47}\\
\;$-$ union \& dense$^{\dagger}$          & 0.25 & 22 & 0.17 & 0.00 & 0.15 & 0.11 & 0.17\\
\;$-$ union \& dense, $-$ text      & --   & -- & 0.17 & 0.00 & 0.00 & 0.06 & 0.11\\
\;$-$ graded evidence (binary $e$)        & 0.50$^{*}$ & not found & 0.42 & 0.27 & 0.38 & 0.36 & 0.47\\
\;$-$ proximity penalty ($\lambda=0$)     & 0.25 & 2  & 0.25 & \textbf{0.36} & 0.23 & 0.28 & \textbf{0.47}\\
DANTE~\cite{luu2025integrated}, same pool & 0.50 & n.e. & \textbf{0.42} & 0.18 & 0.00 & 0.19 & 0.19\\
\midrule
$\beta=0.1$                               & -- & -- & 0.42 & 0.27 & 0.31 & 0.33 & 0.44\\
$\beta=0.5$                               & -- & -- & 0.42 & 0.27 & 0.38 & 0.36 & 0.47\\
$\lambda=0.0025$                          & -- & -- & 0.42 & 0.27 & 0.38 & 0.36 & 0.47\\
$\lambda=0.01$                            & -- & -- & 0.08 & 0.18 & 0.31 & 0.19 & 0.36\\
\midrule
TEMPO-P, keyword queries                  & -- & -- & 0.25 & 0.27 & 0.38 & 0.31 & 0.42\\
DANTE, keyword queries                    & -- & -- & 0.25 & 0.09 & 0.00 & 0.11 & 0.11\\
\bottomrule
\multicolumn{8}{@{}l}{\scriptsize $^{\dagger}$all events required in the top-$M$ of one video;\quad $^{*}$Q1--Q4 carry no textual cue.}
\end{tabular}
\end{table}
\paragraph{\textbf{Results and discussion.}} Table~\ref{tab:diag} shows that TEMPO-P
matches DANTE at R@1 on visual queries; its higher R@10 reflects ranked alternatives
rather than better single-path alignment. DANTE's failure on text-defined events
is a representational limitation, so alignment quality is compared only on visual
queries. Candidate union and dense scoring provide the largest gains, while the
compactness penalty helps visual queries but can hurt OCR-cue queries. Defaults
lie on a performance plateau, although excessive $\lambda$ degrades retrieval;
graded evidence is supported only by Q5 and remains inconclusive. Full-sentence
queries outperform keywords mainly on visual events, with text-defined events
unaffected. Conditional on finding the target video, TEMPO-P localizes 73\% of
events within $\pm30$ frames (DANTE 65\%). Dense scoring is the main cost: 50.5\,s per query on
CPU (DANTE 42.3\,s), reduced to under 10\,s on a laptop GPU (NVIDIA RTX~4050), which
is compatible with interactive use. Decoding raw frames
adds latency without improving retrieval, supporting the keyframe-only design.

%% file: secs/conclusion.tex
\section{Conclusion}
We presented EVISTA and TEMPO-P for multimodal and ordered multi-event video
retrieval in the Ho Chi Minh City AI Challenge 2026. TEMPO-P combines candidate
videos across events and modalities, integrates visual and OCR/ASR evidence, and
uses dynamic programming to select temporally ordered, compact keyframe tuples.
On organizer queries and 36 additional queries, it matches visual-only alignment
on visually described events while supporting events defined solely by on-screen
text. Full-sentence descriptions also outperform keyword-only queries with a
multilingual encoder.

\paragraph{\textbf{Limitations and Future Works.}} The 36 additional queries were written by one author
after inspecting the keyframes, with OCR cues verified to be recognized, so the results
on text-defined events are an upper bound under correct recognition; misread, missing, or
undiacritised text is not covered. The weights $\beta$, $\lambda$, and the graded
evidence values were fixed before the evaluation and not tuned. Future work
will use larger, independently annotated query sets, including OCR/ASR failures.
Dense scoring dominates the query time: about 50\,s per query on CPU, reduced to
under 10\,s on a laptop GPU (NVIDIA RTX~4050); caching keyframe scores across related
queries could reduce it further. The evidence weights and the compactness penalty could
also be learned from interaction logs instead of being fixed by hand.

%% file: secs/ack.tex
\section{Acknowledgement}
We sincerely thank Ho Chi Minh City University of Technology -- VNU--HCM for supporting this research. We also extend our gratitude to the HCMUT EE Machine Learning \& IoT Lab for supporting the team's experimental work. 

%% file: refs.bib
@article{lokovc2023interactive,
  title={Interactive video retrieval in the age of effective joint embedding deep models: lessons from the 11th VBS: J. Loko{\v{c}} et al.},
  author={Loko{\v{c}}, Jakub and Andreadis, Stelios and Bailer, Werner and Duane, Aaron and Gurrin, Cathal and Ma, Zhixin and Messina, Nicola and Nguyen, Thao-Nhu and Pe{\v{s}}ka, Ladislav and Rossetto, Luca and others},
  journal={Multimedia Systems},
  volume={29},
  number={6},
  pages={3481--3504},
  year={2023},
  publisher={Springer}
}

@inproceedings{do2023news,
  title={News event retrieval from large video collection in Ho Chi Minh City AI challenge 2023},
  author={Do, Trong-Le and Nguyen, Hai-Dang and Nguyen, Quang-Thuc and Tran, Mai-Khiem and Huynh, Viet-Tham and Gurrin, Cathal and Ninh, Tu V and Le, Tu-Khiem and Ngo, Thanh Duc and Ngo, Tu-Trinh and others},
  booktitle={Proceedings of the 12th International Symposium on Information and Communication Technology},
  pages={1011--1017},
  year={2023}
}

@inproceedings{do2025toward,
  title={Toward abstraction-level event retrieval in large video collections: leveraging human knowledge and LLM-based reasoning in the Ho Chi Minh City AI Challenge 2025},
  author={Do, Trong-Le and Huynh, Viet-Tham and Nguyen, Hai-Dang and Nguyen-Quang, Thuc and Tran, Mai-Khiem and Nguyen, Trong-Thuan and Ninh, Tu V and Le, Tu-Khiem and Ngo, Thanh Duc and Dang-Nguyen, Duc-Tien and others},
  booktitle={International Symposium on Information and Communication Technology},
  pages={124--138},
  year={2025},
  organization={Springer}
}

@inproceedings{nguyen2025vidalign,
  title={VidAlign: Integrating Multi-event Alignment and LLM Co-searching for Video Retrieval},
  author={Nguyen, Huu-Tuan and Tran, Dien X and Vo, Minh-Thinh and Nguyen, Thanh Nhan and Nguyen Thai, Thanh-Long and Le, Phuc-Lu},
  booktitle={International Symposium on Information and Communication Technology},
  pages={178--188},
  year={2025},
  organization={Springer}
}

@inproceedings{radford2021learning,
  title={Learning transferable visual models from natural language supervision},
  author={Radford, Alec and Kim, Jong Wook and Hallacy, Chris and Ramesh, Aditya and Goh, Gabriel and Agarwal, Sandhini and Sastry, Girish and Askell, Amanda and Mishkin, Pamela and Clark, Jack and others},
  booktitle={International conference on machine learning},
  pages={8748--8763},
  year={2021},
  organization={PmLR}
}

@inproceedings{wang2023image,
  title={Image as a foreign language: Beit pretraining for vision and vision-language tasks},
  author={Wang, Wenhui and Bao, Hangbo and Dong, Li and Bjorck, Johan and Peng, Zhiliang and Liu, Qiang and Aggarwal, Kriti and Mohammed, Owais Khan and Singhal, Saksham and Som, Subhojit and others},
  booktitle={2023 IEEE/CVF Conference on Computer Vision and Pattern Recognition (CVPR)},
  pages={19175--19186},
  year={2023},
  organization={IEEE}
}

@inproceedings{amato2024visione,
  title={Visione 5.0: Enhanced user interface and ai models for vbs2024},
  author={Amato, Giuseppe and Bolettieri, Paolo and Carrara, Fabio and Falchi, Fabrizio and Gennaro, Claudio and Messina, Nicola and Vadicamo, Lucia and Vairo, Claudio},
  booktitle={International Conference on Multimedia Modeling},
  pages={332--339},
  year={2024},
  organization={Springer}
}

@inproceedings{heller2022multi,
  title={Multi-modal interactive video retrieval with temporal queries},
  author={Heller, Silvan and Arnold, Rahel and Gasser, Ralph and Gsteiger, Viktor and Parian-Scherb, Mahnaz and Rossetto, Luca and Sauter, Loris and Spiess, Florian and Schuldt, Heiko},
  booktitle={International Conference on Multimedia Modeling},
  pages={493--498},
  year={2022},
  organization={Springer}
}

@inproceedings{ma2024leveraging,
  title={Leveraging LLMs and generative models for interactive known-item video search},
  author={Ma, Zhixin and Wu, Jiaxin and Ngo, Chong Wah},
  booktitle={International Conference on Multimedia Modeling},
  pages={380--386},
  year={2024},
  organization={Springer}
}

@inproceedings{cormack2009reciprocal,
  title={Reciprocal rank fusion outperforms condorcet and individual rank learning methods},
  author={Cormack, Gordon V and Clarke, Charles LA and Buettcher, Stefan},
  booktitle={Proceedings of the 32nd international ACM SIGIR conference on Research and development in information retrieval},
  pages={758--759},
  year={2009}
}

@article{sakoe1978dynamic,
  title={Dynamic programming algorithm optimization for spoken word recognition},
  author={Sakoe, Hiroaki and Chiba, Seibi},
  journal={IEEE transactions on acoustics, speech, and signal processing},
  volume={26},
  number={1},
  pages={43--49},
  year={1978},
  publisher={IEEE}
}

@inproceedings{zhukov2019cross,
  title={Cross-task weakly supervised learning from instructional videos},
  author={Zhukov, Dimitri and Alayrac, Jean-Baptiste and Cinbis, Ramazan Gokberk and Fouhey, David and Laptev, Ivan and Sivic, Josef},
  booktitle={2019 IEEE/CVF Conference on Computer Vision and Pattern Recognition (CVPR)},
  pages={3532--3540},
  year={2019},
  organization={IEEE}
}

@article{dvornik2021drop,
  title={Drop-dtw: Aligning common signal between sequences while dropping outliers},
  author={Dvornik, Mikita and Hadji, Isma and Derpanis, Konstantinos G and Garg, Animesh and Jepson, Allan},
  journal={Advances in Neural Information Processing Systems},
  volume={34},
  pages={13782--13793},
  year={2021}
}

@inproceedings{han2022temporal,
  title={Temporal alignment networks for long-term video},
  author={Han, Tengda and Xie, Weidi and Zisserman, Andrew},
  booktitle={2022 IEEE/CVF Conference on Computer Vision and Pattern Recognition (CVPR)},
  pages={2896--2906},
  year={2022},
  organization={IEEE}
}

@inproceedings{vo2025enhanced,
  title={Enhanced multimodal video retrieval system: Integrating query expansion and cross-modal temporal event retrieval},
  author={Vo, Van-Thinh and Nguyen, Minh-Khoi and Tran, Minh-Huy and Nguyen-Tran, Anh-Quan and Nguyen, Duy-Tan and Nguyen Khanh, Loi and Phan, Anh-Minh},
  booktitle={International Symposium on Information and Communication Technology},
  pages={166--177},
  year={2025},
  organization={Springer}
}

@inproceedings{luu2025integrated,
  title={Integrated Semantic and Temporal Alignment for Interactive Video Retrieval},
  author={Luu, Thanh-Danh and Nguyen-Dinh, Le-Vu and Bui, Duy-Bao and Tran, Duc-Thien and Le, Nam-Tien and Nguyen-Nhu, Tinh-Anh},
  booktitle={International Symposium on Information and Communication Technology},
  pages={96--107},
  year={2025},
  organization={Springer}
}

@inproceedings{vu2027madtempo,
  title={MADTempo: an interactive system for multi-event temporal video retrieval with query augmentation},
  author={Vu, Huu-An and Mai, Van-Khanh and Nguyen, Trong-Tam and Dam, Quang-Duc and Nguyen, Tien-Huy and Le, Thanh-Huong},
  booktitle={International Symposium on Information and Communication Technology},
  pages={179--193},
  year={2027},
  organization={Springer}
}

@inproceedings{tran2025towards,
  title={Towards efficient and robust moment retrieval system: A unified framework for multi-granularity models and temporal reranking},
  author={Tran, Huu-Loc and Nguyen-Nhu, Tinh-Anh and Phan-Nguyen, Huu-Phong and Nguyen, Tien-Huy and Nguyen-Dich, Nhat-Minh and Dao, Anh and Do, Huy-Duc and Nguyen, Quan and Le, Hoang M and Dinh, Quang-Vinh},
  booktitle={2025 IEEE/CVF Conference on Computer Vision and Pattern Recognition Workshops (CVPRW)},
  pages={3719--3729},
  year={2025},
  organization={IEEE}
}

@inproceedings{nguyen2025lightweight,
  title={A lightweight moment retrieval system with global re-ranking and robust adaptive bidirectional temporal search},
  author={Nguyen-Nhu, Tinh-Anh and Tran, Huu-Loc and Le, Nguyen-Khang and Nguyen, Minh-Nhat and Nguyen, Tien-Huy and Nguyen-Huu, Hoang-Long and Phan-Nguyen, Huu-Phong and Pham, Huy-Thach and Nguyen, Quan and Le, Hoang M and others},
  booktitle={2025 IEEE/CVF Conference on Computer Vision and Pattern Recognition Workshops (CVPRW)},
  pages={3708--3718},
  year={2025},
  organization={IEEE}
}

@article{yang2025qwen3,
  title={Qwen3 technical report},
  author={Yang, An and Li, Anfeng and Yang, Baosong and Zhang, Beichen and Hui, Binyuan and Zheng, Bo and Yu, Bowen and Gao, Chang and Huang, Chengen and Lv, Chenxu and others},
  journal={arXiv preprint arXiv:2505.09388},
  year={2025}
}

@article{bolya2025PerceptionEncoder,
  title={Perception Encoder: The best visual embeddings are not at the output of the network},
  author={Daniel Bolya and Po-Yao Huang and Peize Sun and Jang Hyun Cho and Andrea Madotto and Chen Wei and Tengyu Ma and Jiale Zhi and Jathushan Rajasegaran and Hanoona Rasheed and Junke Wang and Marco Monteiro and Hu Xu and Shiyu Dong and Nikhila Ravi and Daniel Li and Piotr Doll{\'a}r and Christoph Feichtenhofer},
  journal={arXiv},
  year={2025}
}

@misc{easyocr2024,
  author       = {JaidedAI},
  title        = {{EasyOCR}: Ready-to-use OCR with 80+ supported languages},
  year         = {2024},
  publisher    = {GitHub},
  journal      = {GitHub repository},
  howpublished = {\url{https://github.com/JaidedAI/EasyOCR}},
  note         = {Version 1.7.2}
}

@misc{vietocr2020,
  author       = {Pham, B. Quoc},
  title        = {{VietOCR}: Transformer OCR for Vietnamese Text Recognition},
  year         = {2020},
  publisher    = {GitHub},
  journal      = {GitHub repository},
  howpublished = {\url{https://github.com/pbcquoc/vietocr}},
  note         = {Accessed: 2026-10-04}
}

@inproceedings{PhoWhisper,
  title     = {{PhoWhisper: Automatic Speech Recognition for Vietnamese}},
  author    = {Thanh-Thien Le and Linh The Nguyen and Dat Quoc Nguyen},
  booktitle = {Proceedings of the ICLR 2024 Tiny Papers track},
  year      = {2024}
}

@misc{yolo11_ultralytics,
      author = {Glenn Jocher and Jing Qiu},
      title = {Ultralytics YOLO11},
      version = {11.0.0},
      year = {2024},
      url = {https://github.com/ultralytics/ultralytics},
      orcid = {0000-0001-5950-6979, 0000-0003-3783-7069},
      license = {AGPL-3.0}
    }

@article{honicke2026jina,
  title={jina-embeddings-v5-omni: Geometry-preserving Embeddings via Locked Aligned Towers},
  author={H{\"o}nicke, Florian and G{\"u}nther, Michael and Koukounas, Andreas and Akram, Mohammad Kalim and Martens, Scott and Sturua, Saba and Xiao, Han},
  journal={arXiv preprint arXiv:2605.08384},
  year={2026}
}
